\documentstyle[epsfig]{aipproc}

\begin{document}
\title{BATSE Sky Exposure} 
\author{Jon Hakkila$^*$, Charles A. Meegan$^{\ddagger}$, Geoffrey N. 
Pendleton$^{\dagger}$, William Henze$^{\circ}$, Michael McCollough$^{\star}$,\\
Jefferson M. Kommers$^{\diamond}$, and Michael S. Briggs$^{\dagger}$}

\address{$^*$Mankato State University, Mankato MN  56002-8400 \\
$^{\ddagger}$NASA/Marshall Space Flight Center, Huntsville, AL 35824 \\
$^{\dagger}$University of Alabama in Huntsville, Huntsville, AL 35899 \\
$^{\circ}$Teledyne Brown at NASA/MSFC, Huntsville, AL  35824 \\
$^{\star}$USRA at NASA/MSFC, Huntsville, AL 35824 \\
$^{\diamond}$Massachusetts Institute of Technology, Cambridge, MA  02139}

\maketitle

\begin{abstract}
Angular sky exposure is presented for a number of published BATSE 
gamma-ray burst catalogs. A new algorithm was required due to telemetry
gaps resulting from onboard tape recorder failures; the new algorithm
improves the 1B Catalog exposure calculation. The most influential
effects limiting BATSE's exposure are (1) deadtime due to triggering, (2) sky
blockage by the Earth, and (3) trigger disabling when the spacecraft is in the
SAA and over other specific Earth locations. Exposure has improved
during the CGRO mission as a result of decreased Solar flares and
magnetospheric particle events.
\end{abstract}

\section*{Introduction}
Analyses of the angular and intensity distributions of gamma-ray bursts require
knowledge of instrumental sensitivities and biases (e.g. 
\cite{Higdon90,Hakkila94,Loredo95,Briggs96}). It is important to have 
available not just a catalog of gamma-ray bursts detected by
an experiment, but a summary describing the sensitivity of the experiment to
detecting gamma-ray bursts. The sensitivity is generally subdivided into two
components: angular exposure (describing the angular instrumental
sensitivity), and trigger efficiency (describing the flux-dependent
instrumental sensitivity). Angular exposure is described here; the
algorithm to be used for calculating
trigger efficiency is described elsewhere \cite{Pendleton98}.

The angular exposure was initially calculated for the BATSE 1B Catalog
\cite{Brock92}, but was not updated because onboard flight recorder failures 
produced telemetry gaps in the data. A statistical approach was needed rather 
than the previous procedure of direct extraction of the exposure information. 
The 1B analysis helped in designing this approach, however, as a number of 
important considerations were identified:

\begin{itemize}
\item Zeroth order exposure is due to experimental livetime (burst 
detection is impossible if high voltage is off or if onboard triggers 
are disabled).
\item The first order exposure effect is sky blockage by the Earth, which 
introduces a quadrupole moment in the coverage such that bursts are 
less-easily detected in the equatorial plane.
\item Second order exposure effects are a dipole moment and quadrupole moments
altered by the SAA and trigger disable boxes.
\end{itemize}

Small right ascension-dependent exposure components result from CGRO's 
52-day precession period, but these components have been found to be small
for BATSE catalogs since these span more than one precession cycle. Other
higher-order effects have also been omitted because they are small.

\section*{Analysis}

Calculation of the angular exposure proceeds as follows:

\begin{itemize}
\item A CGRO orbital model is used to step the satellite through 30-second
positional increments during the BATSE catalog in question. Inclusive
dates of specific BATSE catalogs are given in Table \ref{table1}.

\begin{table}
\caption{BATSE catalog start and end times.}
\label{table1}
\begin{tabular}{||c|c|c||}
Catalog & start time (tjd) & stop time (tjd) \\
\tableline
1B catalog & 8358 & 8686 \\
2B-1B catalog & 8686 & 9055 \\
3B-2B catalog & 9055 & 9614 \\
4B-3B catalog & 9614 & 10324 \\
3B catalog & 8358 & 9614 \\
4B catalog & 8358 & 10324 \\
\end{tabular}
\end{table}

\item At the beginning of each day (labeled by tjd, or Truncated Julian Date), 
parameters relating to BATSE operations are
checked and loaded. These parameters include trigger disable boxes,
number of triggers occurring, and mean trigger durations. At each time step,
the satellite is either found to be in the SAA or trigger disable box (in which
case it is unable to trigger) or outside of these. The amount of livetime (time
spent outside trigger disable boxes) is thus strongly dependent on satellite
latitude.

\item The Earth and its atmosphere block a substantial portion of the sky from 
BATSE at any given time; averaged over many orbits this produces a 
declination dependence. The Earth center declination and satellite altitude are
used to calculate Earth blockage, which is then
used to obtain the declination-dependent livetime.

\item The deadtime due to trigger enabling (originally 90 minutes per trigger,
irrespective of the actual event duration) has been made much more complicated
by the failure of the onboard flight recorders. Telemetry gaps resulting from
these failures have required a number of flight software revisions, and trigger
durations have been recalculated whenever these have been made, and/or when
trigger criteria have changed. The mean trigger duration is composed of a
weighted average of long and short readouts, with longer readouts being
returned when bright triggers occur. The mean trigger durations are shown in
Table \ref{table2}.

\begin{table}
\caption{Mean trigger durations during the BATSE operation time.}
\label{table2}
\begin{tabular}{||c|c|c||}
start time (tjd)& stop time (tjd)& mean trigger duration (sec) \\
\tableline
8358 & 8973 & 5631 \\
8973 & 8995 & 2747 \\
8995 & 9078 & 2177 \\
9078 & 9320 & 1927 \\
9320 & 9495 & 3905 \\
9495 & 9614 & 4854 \\
9614 & 9922 & 5180 \\
9922 & 10092 & 5131 \\
10092 & 10095 & 1721 \\
10095 & 10122 & 2700 \\
10122 & 10182 & 4109 \\
10182 & 10465 & 4741 \\
10465 & 10505 & 4245 \\
10505 & 10617 & 5231 
\end{tabular}
\end{table}

{\em Since trigger efficiency is difficult to calculate during burst 
overwrites, all overwriting bursts must 
be excluded from analysis when using the sky exposure
calculated here in conjunction with the BATSE burst catalogs.}

\item The resulting trigger livetime per tjd $l$ can be calculated 
statistically from the trigger rate $t$ (triggers per tjd), and is

\begin{equation}
l = \left(\frac{d}{86400}\right)^{t}
\end{equation}

where $d$ is the mean trigger duration (in seconds).

\item The overall exposure is the fraction of time that a given location on 
the sky can be observed, given the trigger deadtime, the Earth blockage, and 
the orbital trigger disabling.
\end{itemize}

\section*{Results}

\begin{figure}[b!] 
\centerline{\psfig{file=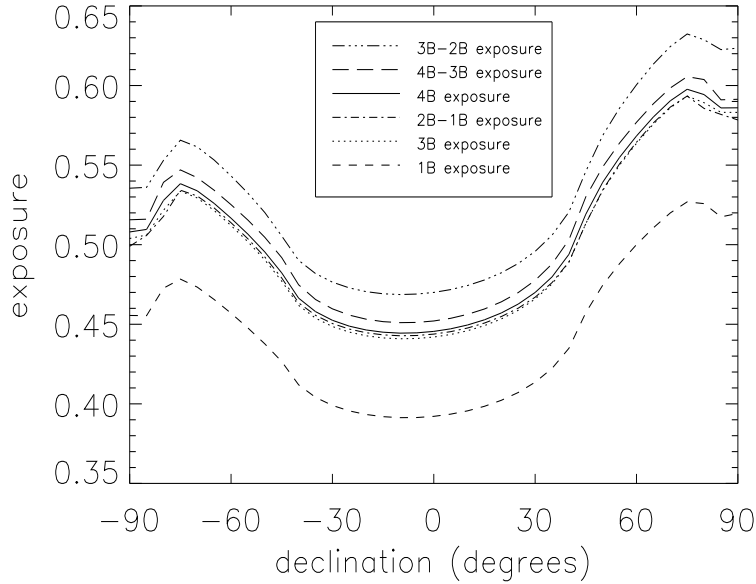,height=3.0in,width=3.8in}}
\vspace{12pt}
\caption{Sky exposure as a function of declination for various BATSE burst
catalogs.}
\label{fig1}
\end{figure}

BATSE's angular exposure is shown in Figure \ref{fig1} for the 
published BATSE
burst catalogs and subcatalogs. Statistical properties of the exposure
are summarized in Table \ref{table3}.

\begin{table}
\caption{Mean exposure, exposure dipole moments, and exposure quadrupole
moments for BATSE catalogs and subcatalogs.}
\label{table3}
\begin{tabular}{||c|c|c|c|c|c||}
Catalog & mean exposure & $\langle \sin \delta \rangle$ & $ \langle \cos \theta
\rangle$ & $\langle \sin^{\rm 2} \delta - 1/3 \rangle$ & $\langle \sin^{\rm 2}
b - 1/3 \rangle $ \\
\tableline
1B & 0.425 & 0.017 & $-$0.008 & 0.025 & $-$0.004 \\
2B-1B & 0.478 & 0.018 & $-$0.009 & 0.024 & $-$0.004 \\
3B-2B & 0.507 & 0.019 & $-$0.009 & 0.024 & $-$0.004 \\
4B-3B & 0.489 & 0.018 & $-$0.009 & 0.024 & $-$0.004 \\
3B & 0.477 & 0.018 & $-$0.009 & 0.024 & $-$0.004 \\
4B & 0.481 & 0.018 & $-$0.009 & 0.024 & $-$0.004 \\
\end{tabular}
\end{table}

There are a number of differences between published 1B results \cite{Brock92} 
and the 1B results presented here. These results affect many
subsequently-published BATSE exposure estimates which were based on the
1B exposure. The most significant differences are as follows:

\begin{itemize}
\item The original 1B calculation considered livetime to include only times 
when the trigger threshold was $5.5\sigma$ in energy channels $2+3$. This left 
as deadtime intervals when the trigger threshold was $5\sigma$ and $10\sigma$. 
By including these other triggering times in the new calculation, the overall 
1B exposure (0.425) is greater than was previously estimated (0.355).

\item The original 1B calculation sampled the sky only on 5 minute intervals.
By not interpolating the spacecraft position, the angular size of the SAA box 
was overestimated, producing a larger dipole moment
($\langle \sin \delta \rangle = 0.026$ and $ \langle \cos \theta
\rangle = -0.013$) and more pronounced quadrupole moments ($\langle 
\sin^{\rm 2} \delta - 1/3 \rangle = 0.026$ and $\langle \sin^{\rm 2}
b - 1/3 \rangle = -0.005$) than those found via the new algorithm.
\end{itemize}

We have compared the results of the new algorithm over one 52-day precession
cycle to that calculated from the untriggered burst search \cite{Kommers98}. 
The untriggered burst
search uses different triggering criteria, no trigger disable boxes, and a
lower detection threshold than BATSE's onboard triggers. Although the general
exposure properties identified by the new BATSE exposure algorithm and the
untriggered burst search are in agreement, there are significant (but easily
understood) differences between them:

\begin{itemize}
\item The overall angular exposure found from the untriggered burst search is 
larger than that found using the onboard BATSE triggers because the lack of 
trigger disable boxes produces more livetime. 

\item The untriggered burst search has significantly larger dipole and 
quadrupole moments than those found using the onboard BATSE triggers, which 
again results from the lack of trigger disable boxes. There have 
been significantly more particle events detected while the spacecraft was over 
the southern hemisphere \cite{Horack92}, so that an experiment such as the
untriggered burst search is more sensitive than the triggered BATSE experiment
towards southern hemisphere observations. 
\end{itemize}

Estimated uncertainties in the exposure fractions are less than 4\%.
Possible sources of systematic error in the moments are currently under study, 
but are expected to be less than 0.003 for the Galactic moments.

\end{document}